# CEP TELEFONU SATIN ALMA KARARINDA MARKANIN ETKİSİ: Y KUŞAĞI TÜKETİCİLER ÜZERİNE BİR ARAŞTIRMA[1]

# THE EFFECT OF THE BRAND IN THE DECISION TO PURCHASE THE MOBILE PHONE: A RESEARCH ON Y GENERATION CONSUMERS

İbrahim Halil EFENDİOĞLU[2]

Adnan Talha MUTLU[3]

Yakup DURMAZ[4]

**Öz**

*Bu çalışmanın amacı, cep telefonu tercihinde markanın satın alma kararına etkisinin Y kuşağı üzerinde belirlemektir. Bu amaç doğrultusunda Y kuşağı yaş aralığında olan ve son bir yılda cep telefonu satın almış 231 kişi ile yüz yüze anket yapılmıştır. Çalışma, Harran Üniversitesi Meslek Yüksek Okullarında görevli genç akademisyenler ve üniversite öğrencileri ile gerçekleştirilmiştir. Toplanan verilerin yapısal eşitlik modellemesi kullanılarak, AMOS ve SPSS istatistiksel paket programları ile analiz edilmiştir. Araştırma sonuçlarına göre markanın prestiji, ismi ve güvenilirliği Y kuşağı tüketicilerin satın kararına önemli ölçüde etki etmektedir. Bunun yanında markanın fiyatı, reklam kampanyaları ve teknik servis hizmetleri de bu kararı daha az etkilemektedir. Diğer taraftan cep telefonu markalarının üretim yaptığı yer ve sosyal sorumluluk projelerine verdiği önem satın alma kararına etki etmemektedir.*

***Anahtar Kelimeler:*** *Y kuşağı, Marka tercihi, Satın alma kararı.*

***Jel Kodu:*** *M31, M39*

**Abstract**

*The aim of the study is to determine the effect of the brand on purchasing decision on generation Y. For this purpose, a face-to-face survey was conducted with 231 people in the Y age range who have purchased mobile phones in the last year. The study was conducted with young academicians and university students working in Harran University Vocational High Schools. The collected data were analysed with AMOS and SPSS statistical package programs using structural equation modelling. According to the results of the research, the prestige, name and reliability of the brand have a significant impact on the purchase decision of the Y generation. Price, advertising campaigns and technical services of the brand also affect this decision less. On the other hand, the place where mobile phone brands are produced and the importance they attach to social responsibility projects do not affect the purchasing decision.*

***Keywords:*** *Generation Y, Brand preference, Purchasing decision.*

***Jel Codes:*** *M31, M39*

---

Makale Türü: Araştırma Makalesi                                    Research Type: Research Paper

## 1.GİRİŞ

Küresel pazar payını verilerine göre 2019 yılının son çeyreğinde tüketiciler dünya genelindeki cep telefonu markaları arasından %20 Apple'ı ,%18,8 Samsung'u, %15,2 Huawei'i, , %8,9 Xiaomi'i, %8,3 Oppo'yu ve %28,7'si diğer markaları tercih etmiştir (Statista, 2019). Dünyanın en değerli markası olan Apple'ın marka değeri bir önceki yıla göre % 12'lik bir artışla 205 milyar dolara yükselirken dünyanın en değerli yedinci markası olan Samsung'un marka değeri % 11'lik bir artışla 53 milyar dolara yükselmiştir (Forbes, 2019). Dünya çapındaki akıllı telefon satıcıları 2019'un son çeyreğinde toplam 369.8 milyon adet cep telefonu satılmıştır (IDC, 2019). Bu durum cep telefonu piyasasında markalar arasında ciddi bir rekabet olduğunu göstermektedir.

Son derece rekabetçi olan cep telefonu pazarında cep telefonu üreticileri, tüketicileri rakiplerinin yerine kendi markalarını seçmeye ikna etmek için sürekli farklılaştırıcı unsurlar bulmaya çalışmaktadır (Das, 2012: 69). Ürünlerin, marka gibi ayırt edici bir özelliğe sahip olmalarının temel nedeni, markalanmış ürünlerin diğer ürünlerden farklılaşmasıdır. Bu durum, markanın temel görevini ifade eder (Öztuğ, 2003: 14). Bu temel görev ise markanın en önemli özelliğini ortaya koymaktadır. Bir ürünün markalanıp, markalanmayacağı veya ne tür bir marka ile pazara sunulacağı konusu işletmeler için düşündürücü bir durumdur. Çünkü ürün, markası ile bütünleştikten sonra marka ürünün ayrılmaz bir parçası olmaktadır. Böylelikle marka, ürüne ait pazarlama stratejileri ve satış hacminin arttırılması için yönlendirici bir etkiye kavuşmaktadır (Türk, 2004: 58). Markanın ürün üzerindeki bu etkisi, ürünün pazarda tutunma ve başarı sağlaması açısından son derece önemlidir.

Toplumda teknoloji odaklı ürünleri daha yaygın olarak kullanan gençler, son yıllarda cep telefonu markalarının ilgi odağı haline gelmiştir. Teknolojiye yön veren bu genç tüketiciler ürün tercihi yaparken marka faktörüne önem vermektedir. Bu doğrultuda bu tercihi dikkate alan markalar genç tüketicilerin cep telefonu marka seçim kararlarında etkili olan faktörleri dikkatle incelemektedir (Shahzad ve Sobia, 2013: 370). Markalar, tüketicilerin günlük yaşamında önemli bir rol oynamaktadır. Markalar ve tüketiciler arasındaki bu yakın ilişki markalı ürünlerin tüketici yaşam tarzının bir unsuru olmasından kaynaklanmaktadır. Bu yüzden işletmeler marka değerini artırmak için tüketicilerin marka tercih sürecindeki bilgisini genişletmesi gerekmektedir (Duarte ve Raposo, 2010: 465). Diğer taraftan marka tercihinin önemi, bilişsel temel tüketicinin hafızasından silindikten sonra bile, duyuşsal bileşenin varlığını sürdürebildiğine işaret ettiğinde tarafından vurgulanmaktadır (Zajonc, 1980). Marka tercihi konusunda yapılan araştırmaların çoğunda farklı faktörler tüketiciler açısından belirleyicidir. Çalışmalara göre tüketicilerin belirli bir markayı seçmesinde hayati derecede önemli olan kilit unsurlar bulunmaktadır (Bushman,1993; Dunn and Murphy 1986; Sengupta and Fitzsimons 2000).

Markalı ürünlere önem veren Y kuşağı tüketiciler, yaklaşık olarak 1980 ile 2000 yılları arasında doğanları kapsamaktadır. Bu kuşak satın alma gücüne sahip genç tüketicilerden oluşmaktadır. Ayrıca bu kuşağa mensup tüketiciler kullandığı ve satın aldığı ürünlerin markalarına önem vermekte ve dikkat etmektedir. Bu doğrultuda çalışmanın amacı Y kuşağı tüketicilerin cep telefonu satın alma kararı verirken markanın hangi özelliğinden etkilendiklerinin belirlenmesidir. Çalışmada markanın fiyatı, reklamı, üretim yeri, ismi, güvenilirliği, prestiji, teknik servis hizmetleri ve sosyal sorumluluk projelerine verdiği önem özellikleri incelenecektir.

Çalışmada öncelikle konuyla ilgili önemli kavramlar açıklanacak ve bu konuda literatürdeki daha önceki araştırmalar irdelenecektir. Ardından araştırma modeli, hipotezler ve bunları destekleyen argümanlar sunulacaktır. Yöntem bölümünde, veri toplama ve analiz süreci hakkında bilgi verilecektir. Son bölümde ise çalışmanın sonucu, teorik ve pratik katkıları tartışılacaktır.

## 2. LİTERATÜR İNCELEMESİ

Gülmez (2005) 415 üniversite öğrencisi ile Sivas Cumhuriyet Üniversitesi ve Tokat Gaziosmanpaşa Üniversitesi'nde yaptığı çalışmada öğrencilerin cep telefonu satın alırken marka tercihlerini incelemiştir. Çalışmasında öğrencilerin maddi özgürlüklerinin henüz bulunmaması sebebiyle seçimi daha çok ailelerinin belirlediği görülmüştür. Ayrıca marka seçiminde kendilerinin de etkin rollerinin yanı sıra kardeşlerinin, arkadaşlarının, akranlarının ve ünlü sanatçıların kullanmış oldukları markaların da etkili olabildiği görülmüştür. Yang, He ve Lee (2007)'nin ABD ve Çin'de yaptığı araştırmaya her bir ülkeden 200 katılımcının cep telefonu satın alma davranışları incelenmiştir. Araştırma sonucunda ABD'li tüketicilerin satın alma aşamasında ürünün fiyatını önemsedikleri, referans grupların ve sosyal çevrenin söylemlerini pek önemsemedikleri belirlenmiştir. Çin'li tüketicilerin ise satın alacakları cihazın tasarımına ve başkalarını etkileme gücüne daha çok önem verdikleri görülmüştür. Maltepe (2009) ABD'deki üniversitelerde eğitim gören 58 öğrenci ve Türkiye'de eğitim göre 119 öğrenci ile toplam 177 kişiyi kapsayan çalışmasında cep telefonu satın alma kararlarını etkileyen faktörlerin neler olduğunu incelemiştir. Çalışmada ayrıca her iki ülke öğrencilerinin cep telefonu satın alma kararları karşılaştırmıştır. Amerikalı öğrencilerin cep telefonu satın almada özellikle eğlence arayışında olmaları önemli etken iken Türk öğrencilerde modayı önemseyen ve cihazın ekstra özelliklere sahip olmasını önemsedikleri görülmüştür. Her iki ülke öğrencilerinin de cep telefonu satın almada ürünün fiyatına ve kullanım özelliklerine de önem verdikleri belirlenmiştir. Şimşek (2011) İzmir ilinde 239 öğrenci ile yaptığı çalışmada tüketicilerin cep telefonu satın alma davranışlarını incelemiştir. Çalışmada tüketicilerin öncelikle marka bilinirliği yüksek ve gösterişli ürünleri tercih ettiği görülmüştür. Bunun yanında tüketicilerin arkadaş tavsiyesini ve internet yorumlarını önem verdikleri görülmüştür. Çemrek ve Filiz (2011) 322 üniversite öğrencisiyle Eskişehir Osmangazi Üniversitesi'nde yaptığı araştırmada öğrencilerin cep telefonu seçim kriterlerinde; fiyat, marka, şarj süresi, kullanım kolaylığı ve boyutunun küçük olması etkenleri araştırılmıştır. Araştırmaya katılan kullanıcıların çoğunluğu markaya, fiyata ve kullanım kolaylığına önem vermiştir. Das (2012) çalışmasında cep telefonlarının fiyatı, özellikleri, marka adı, algılanan kullanım kolaylığı ve sağlamlığı, reklam, önerileri ve cihazın endüstriyel tasarımının ürün tercihine etkisini araştırmışlardır. Sata (2013) Etiyopya'nın Hawassa şehrinde 246 katılımcı ile gerçekleştirdiği bu çalışmada tüketicilerin cep telefonu satın alma kararlarını etkileyen faktörler araştırılmıştır. Çalışma sonucunda tüketicilerin cihaz alımında öncelikli etkenin fiyat olduğu bunu ardında gelen en büyük etkenin de cihaz özelliklerinin satın almada önemli bir motivasyonel rol üstlendiği görülmüştür. Khan ve Rohi (2013) 110 üniversite öğrencisi ile Pakistan'da yaptığı çalışmalarında gençlerin cep telefonu marka seçim kriterlerinde; fiyatın, kalitenin, ürün özelliklerinin, aile ve arkadaş önerilerinin, marka imajının, yenilikçi özelliklerin, reklamların ve bu reklamlarda ünlülerin rol almasının, kullanıcı dostu olmasının, şık görünmesinin ve satın alma sonrası hizmetlerin tercih etmeye etkisini incelemişlerdir. Ayrıca kalitenin, marka imajının, aile ve arkadaş tavsiyelerinin gençlerin marka seçimini etkileyen kilit değişkenler olduğunu bulmuşlardır. Riyath ve Musthafa (2014) 82 üniversite öğrencisiyle Sri Lanka'da yaptıkları çalışmada cep telefonu satın alma kararında markanın etkisini, kalitesini, tanıtım ve reklam özelliklerini, fiyatını ve sosyal çevre etkenlerini incelemişlerdir. Araştırma sonucunda Sri Lanka'da üniversite öğrencileri cep telefonu satın alırken ürünün en fazla fiyat kısmına dikkat ettikleri görülmüştür. Çakır ve Demir (2014) 387 üniversite öğrencisi ile Aydın ilinde yaptığı çalışmada akıllı telefon satın almada ürünün fiyat, marka, sosyal faktörler, reklam faaliyetleri, cihaz özellikleri, beklenen ve algılanan kullanım kolaylığı gibi unsurların tercihe etkisini araştırmışlardır. Bu araştırma sonucunda cep telefonu satın almada fiyat, sosyal faktör, statü gibi etkenlerin çok etkili olmadığı ve tercihi en çok etkileyen unsurun reklam ve markaya olan güven olduğu görülmüştür. Uddin, Zahan ve Oheduzzaman (2014)'nın 160 katılımcı ile Bangladeş'te gerçekleştirdikleri çalışmada tüketicilerin cep telefonu seçiminde etkilendikleri faktörler araştırılmıştır. Çalışma sonucunda tüketicilerin cep telefonu satın almada en önemli faktörün cihazın fiziksel özellikleri olup bunu yanı sıra fiyat, içerik özellikleri, sosyal statü belirleme yetisi, iş arkadaşlarının ve komşularının marka önerisi, cihazın ağırlığı gibi etkenlerin de satın almalarında önemli rol aldıkları görülmüştür. Polat ve Maksudunov (2015) Kırgızistan'da 400 üniversite öğrencisiyle yaptıkları araştırmada öğrencilerin mobil telefon tercihinde etkili olan faktörleri aratırmışlardır. Araştırmaya katılan öğrenciler için konuşma, müzik dinleme, internet, fotoğraf çekme, diğer cihazlara kablosuz bağlanabilme faktörlerinin tercihi etkilediği görülmüştür. Durmaz ve Şerbetçi (2015) 251 öğrenci katılımıyla Hasan Kalyoncu Üniversitesi ve Kahramanmaraş Sütçü İmam Üniversitesi'nde yaptıkları araştırmada öğrencilerin cep telefonu satın almada etki eden faktörleri aratırmışlardır. Araştırma sonucunda cep telefonu satın almada tercihi büyük ölçüde fiyatın etkilediği ancak sosyal çevrede kullanılan markanın da tercihi etkilediği görülmüştür. Mramba (2015) tüketicilerin cep telefonu satın alma kararlarını Tanzanya'da 160 kişi ile incelenmiştir. Çalışma sonucunda tüketicilerin cep telefonu satın alırken cihazın daha çok dayanıklılık, uzun süre kullanılabilme özelliğini, ihtiyaçlarını karşılayabilme yetisini önemsedikleri bunun yan ısıra ürün menşeisini de inceledikleri görülmüştür. Tatlı (2015) 400 öğrenciye Bingöl Üniversitesi'nde yaptığı çalışmada tüketicilerin akıllı telefon tercihlerinde; gelirin, yaşın, cinsiyetin, medeni durumun, cep telefonu kullanımına bakış açısının ve cihazın görselliğinin etkilerini araştırmıştır. Araştırma sonucuna göre cep telefonu satın alırken gelirin, yaşın ve cep telefonu kullanımına bakış açısının önemli derecede etki ettiği görülürken cinsiyet ve medeni durumun tercihe etkisi görülmemiştir. Yavuz ve Karabulut (2016) Erzincan Üniversitesi'nde eğitim gören 520 öğrenciyle gerçekleştirdikleri araştırmada tüketicilerin cep telefonu satın almada marka tercihlerini nasıl gerçekleştirdikleri ve o markaya olan sadakatlerini araştırmışlardır. Araştırma sonucunda cep telefonu satın almada sosyal çevre ve statü etkeni, tercihi önemli derecede etkilediği erkek kullanıcıların marka sadakatlerinin bayan kullanıcılardan daha düşük seviyede olduğu görülmüştür. Akın (2017) Ömer Halisdemir Üniversitesi'nde eğitim gören 416 katılımcı ile gerçekleştirdiği araştırmada cep telefonu sektöründe genç tüketicilerin marka tercihinde memnuniyetlerini ve tüketim sonrası markaya olan sadakatlerini araştırmıştır. Bu araştırma sonucunda katılımcıların kullandıkları markada olan memnuniyetlerinin sadakatleri üzerinde olan hem davranışsal hem de tutumsal boyutta incelemişlerdir. Araştırmada tüketicilerin memnuniyetlerinin davranışsal sadakatte yoğunlaştığı istatistiksel olarak saptanmıştır. Bu da davranışsal sadakat boyutunun tüketicilerin gerek kullanmış oldukları cihazı değiştirmede gerekse başka bir gereksinimlerinde aynı marka ürünü tercih edecekleri anlamına gelmektedir. Kıral (2018) Adana ilinde ikamet eden 750 katılımcı ile cep telefonu kullanımına bağlı olarak artan cep telefonu operatör tercihlerine etki eden unsurları araştırmıştır. Araştırma sonucunda 2017 senesinde en çok tercih edilen GSM operatörü Turkcell olduğu fakat bu operatörün tercih olasılıklarının zaman içerisinde azalacağı ve uzun dönemde Vodafone operatörünün diğer operatörlerin önüne geçeceği ortaya çıkartılmıştır. Durmaz ve Dağ (2018) 400 katılımcı ile Gaziantep ilinde yaptıkları araştırmada tüketicilerin cep telefonu satın alırken tercih etkenlerini ve satın alınan markaya olan sadakatlerini araştırmışlardır. Araştırma sonucunda tüketicilerin ürün satın almalarında tercihlerini ürünün orijinalliği daha çok etkilemekte ve orijinal olan ürünleri daha çok tercih etmektedirler. Aynı zamanda da orijinal ürünlerin markalarına daha çok sadık kaldıkları görülmüştür.

## 2.1. Marka ve Markanın Özellikleri

Tarihte marka kavramının ilk ortaya çıkışı işaretlenmiş hayvanlar olarak bilinir. Amaç hayvanı diğerlerinden ayırmak ve grubunu tanımlamaktır. Aslında bu durum günümüz uygulamalarıyla paralellik göstermektedir. Çünkü genel anlamda marka, başkalarının ürün ve hizmetlerini ayırt edebilmek için ayırt edici bir işarettir. Bu işaretler isim, logo ve sembollerin birleşimidir (Efendioğlu ve Durmaz, 2016: 56). İnsanların kullandıkları bir ürünü beğenip beğenmemeleri, gelecekte bu malı yeniden nasıl temin edebileceği, mal üzerinde bulunan kusurlardan kimleri sorumlu tutacakları bir bakıma marka kavramıyla ilgilidir. Markalama işlemi toplumsal yaşamın bir gerekliliği olarak meydana gelmiştir. Yapılan araştırmalarda Mısırlılar, Romalılar, Çinliler ve Yunanlılar gibi medeniyetlerin sahipliği belirtmek için çanak, çömlek ve diğer eşyalarının üzerlerine mühür vurdukları görülmüştür (Salman ve Uydacı, 2012: 11). Ürün ya da hizmetin tüketici tarafından kolaylıkla tanınması, reklam ve tanıtımlar sayesinde tüketici ile iletişim halinde olunması pazarlama politikaları açısından markanın ne denli önemli olduğunu ortaya çıkarmıştır. Bu anlamda marka, benzer özelliklere sahip ürün ya da hizmetin diğerlerinden farklılaşmasını sağlayan bir araçtır (Aktuğlu, 2008: 11). Bununla birlikte özellikle somut ürünleri birbirinden ayıran marka, tüketicinin satın alma karar sürecinde önemli bir yere sahip olduğundan, pazarlama ve reklam faaliyetlerinin odak noktası haline gelmiştir.

Markanın, ticaret unvanı ile karıştırılmaması gerekir. Ticaret unvanı, her tacirin ticari işletmesine ilişkin iş ve işlemlerinde kullandığı isimdir ve taciri diğer tacirlerden ayırt etmeye yarar. Marka ise; bir teşebbüsün mal ve hizmetlerini diğer teşebbüsün mal ve hizmetlerinden ayırt etmeye yarar (Zengin, 2012). Ticaret unvanı, işletme ile sahibi arasındaki ilişkiyi yani ticari işletmenin sahibini gösterir. Marka ise, piyasaya sürülen mal ile ticari işletme arasındaki ilişkiyi kurar ve o malı piyasadaki benzerlerinden ayırt etmeye yarar. Her tacir, ticaret unvanı seçip kullanmak zorunda olduğu halde; marka seçmek ve kullanmak zorunda değildir. Ürünlerin rakiplerinden ayrılabilmesi için kullanılan her türlü isim, sembol, logo, amblem, terim ya da bunların bir karışımı olarak adlandırılan marka; günümüzde bu yaklaşımın ötesinde bir 'değerler kümesi' anlamı kazanmıştır. Bu değerler kümesi, ürünün sağladığı işlevsel faydaların yanı sıra hedef tüketicinin ürünü satın alırken önem verdiği ve aradığı (markanın yaşam tarzı, kişiliği yansıtma vb.) katma değerlerden oluşmaktadır. Bu nedenle markalama kararları; markanın objektif nitelikleri, vaat ettikleri, kişiliği, kültürü, tüketicilerle bağlantısı, vizyonu, gücü, denkliği, ömrü, maliyeti, ambalajı gibi faktörlerle örtüşecek şekilde gerçekleştirilmelidir (Arslan v.d.,2001). Bu doğrultuda marka kullanmanın, talep yaratarak tutundurmaya yardımcı olma, rakip mallardan kaynaklı satış kaybını engelleme, siparişlerin yerine getirilmesini ve takibini kolaylaştırma gibi yararları vardır (Mucuk, 1999:150). Markanın pazarlama alanındaki anlamı psikolojik ve sembolik olarak ifade edilebilir. Tüketicilerin ürünü nasıl gördüğü, ürün hakkındaki fikirleri, marka imajı ve değeri markanın psikolojik yönünü ifade etmektedir. Günümüzde marka olarak ifade edilen kavram aslında arkanın psikolojik yönüdür. Ürünlerin marka ile isimleştirilmesi işletmeler açısından önemli bir konudur. Bir marka isminin belirlenmesinden sonra, ismini değiştirmek oldukça zordur. İşletmelerin pazarda başarılı olabilmesi için rakiplerinden farklı bir strateji yürütmesi gerekmektedir. Marka değeri, işletmelerin bütün organizasyonel birimlerini ilgilendirmekle beraber, tüketicilerin algısını yönlendirici özelliğinden dolayı pazarlamanın önemli bir elemanıdır (Besen, 2002:37). Marka, işletmelerin ürünlerini rakip firmalardan ayırt etmesi, farklılaştırması için önemli bir etmendir. Bu farklılaştırma amacı doğrultusunda işletmeler reklam ve promosyon stratejilerine yönelmektedir. İşletmelerin imajı marka isimleri etrafında oluşturulur ve tüketicilerin ürünleri tekrarda satın alması sağlanır. Ayrıca reklamlar ve promosyonlar sayesinde işletmeler pazar paylarını yükseltmektedir (Arkan ve Karaçor, 2014:34).

**2.2.Satın Alma Kararı**

Tüketici ihtiyacının oluşması, bu ihtiyacını gidermeye yönelik yapılan araştırmalar, seçenekleri değerlendirme, satın alam kararının verilmesi, satın alınan ürün ya da hizmetin tüketimi ve sonrasındaki davranışları kapsayan süreç satın alma karar süreci olarak ifade edilmektedir (İslamoğlu ve Altunışık, 2008: 7). Esas olarak tüketicinin gösterdiği davranış satın almadan daha önce, satın alma anında ardından satın alma sonrası süreçler olarak üç aşamalıdır. Bunların içinde hedef pazar seçimi, pazar bölümleme, konumlandırma ve uygun pazarlama karmasının geliştirilmesi ile ilgili önemli kararlar bu durumlar ile oluşur. Satın alma karar süreci olarak bilinen bu süreç ise beş kısımdan oluşur. Bunlar; ihtiyacın ortaya çıkması, alternatiflerin belirlenmesi, alternatiflerin değerlendirilmesi, satın alma kararı ve satın alma sonrası davranıştır (Ercis, Ünal ve Can, 2007:283).

Tutum, niyet ve davranış birbirini etkileyen ve genellikle belirli bir düzende oluşan kavramlardır. Planlı Davranış Teorisi'ne göre kişinin davranışlarını yalnızca kişinin iradesiyle gerçekleşmez bunun yanında inanç, tutum, sübjektif norm ve algılanan davranışsal kontrol de kişinin davranışını belirler. Dolayısıyla önce niyetin belirlendiği, buna bağlı olarak da davranışın oluştuğu görülmektedir (Efendioğlu, 2019: 2180). Satın alma karar sürecinin ilk adımı ihtiyacın belirlenmesidir. Tüketiciler ürün veya hizmeti ihtiyaç ve istek duydukları zaman satın alırlar. İhtiyacın ortaya çıkma aşamasında uyarılar içeriden ya da çevreden gelebilir. Kimi zaman bu ihtiyaç ve istekler bir anda oluşabildiği gibi kimi zaman ise tüketicinin bilinçaltına yerleşmiş ve uzun süre sonra ortaya çıkmaktadır. Bilginin toplanması tüketici, ihtiyacını karşılayacak mamul ve markalara ilişkin bilgi toplar. Bilgileri aileden, arkadaşlardan ya da medyadan edinebilir. Satış noktalarını, her bir mamul, hizmet ya da markanın özelliklerini, fiyatını, ödeme koşullarını tüketicinin öğrenebileceği bilgilerdir (Yükselen, 1994: 49). Tüketici alternatifleri değerlendirirken, ürün ya da hizmetin maliyeti, performansı gibi objektif kriterleri ve markanın değeri, imajı gibi sübjektif değerlendirme kriterlerini kullanır. Ürün ya da hizmetin niteliğine göre değerlendirme kriterleri sayısı da değişmektedir. Daha basit ihtiyaçlara yönelik ürün ve hizmet seçiminde kullanılacak kriter sayısı daha az olurken niteliğin ve fiyatın artmasıyla kullanılacak kriter sayısı da daha çok olmaktadır (Vural, 2007: 43). Satın alma sürecinde tüketiciler, satın almak istediği ürün, hizmet veya markayı seçerler. Tüketiciler birçok alternatif arasından değerlendirmeler soncu ihtiyaçlarına en iyi şekilde cevap verecek markayı tercih etme kararı alırlar. Ancak bu karar sonrası hemen satın alam eylemi gerçekleşmeyebilmektedir. Tüketici tercih ettiği markayı ne zaman, nasıl ve hangi yerden satın alacağına karar verir. Satın alma sürecinde tüketiciler belirlediği ürün veya hizmetin yanında aslında satın alma işlemini yapacağı yeri veya kişiyi de belirlemektedir. Satış işlemini yapan kişinin ya da yerin tercihi ürün ya da hizmetin satın alma seçiminde değişiklik oluşturabilir. Çünkü satın alma koşulları satıcılara göre değişiklik gösterebilmektedir (Fırat ve Azmak, 2007: 253). Tüketiciler bir ürün ya da hizmeti satın alma işlemini gerçekleştirdikten sonra ihtiyaçlarına bağlı olarak ya tatmin olacaklardır ya da olmayacaklardır. Ürün, hizmet veya markanın satış işlemi gerçekleştikten sonra işletmelerin görevi sona ermez. İşletmeler, satış sonrasında tüketicinin tatminine yönelik olumlu ya da olumsuz düşüncelerini ve davranışlarını takip etmelidirler (Kotler, 2000: 182).

### 2.3.Y Kuşağı

1980 ile 2000 arasında doğan bireyleri kapsayan Y kuşağı 'Milenyum Kuşağı' olarak da bilinmektedir (Adıgüzel vd., 2014: 173). Ayrıca bu kuşak alışverişe önem veren hatta fazla tüketim ile anılmaktadır. Bunun yanında çevresindekileri tüketim yaparak etkilemekte ve birbirlerinden de etkilenmektedir. Bu kuşak bilgili, sorgulayan, araştıran, bilinçli ve sosyal sorumluluklara dikkat eden bireylerden oluşmaktadır. Bundan dolayı pazarlama alanında araştırmalara sık sık konu olmaktadır (Baycan, 2017: 5). Bu kuşağın mensuplarının markalara ve firmalardan aldıkları hizmete dikkat etmektedir. Üstelik marka ile ilgili bir sorun olduğunu düşündüklerinde bunu arkadaşları arasında hızla yaymaktadır (Yücel Güngör, 2018: 56). Dolayısıyla görünüşlerine, giyimlerine, kullandıkları markalara hassasiyet gösterir ve imajları ile daha çok kişiselleşmek isterler (William ve Page, 2011: 8-10).

### 3.HİPOTEZ GELİŞTİRME

Literatür taraması çerçevesinde geliştirilen ve çalışma kapsamında test edilen araştırma modeli Şekil 1'de yer almaktadır. Bu bölümde tarif edilen araştırma modeliyle ilgili değişkenler ve ilgili literatüre dayalı hipotezler aşağıda sunulmuştur.

**Şekil 1:** Araştırma Modeli

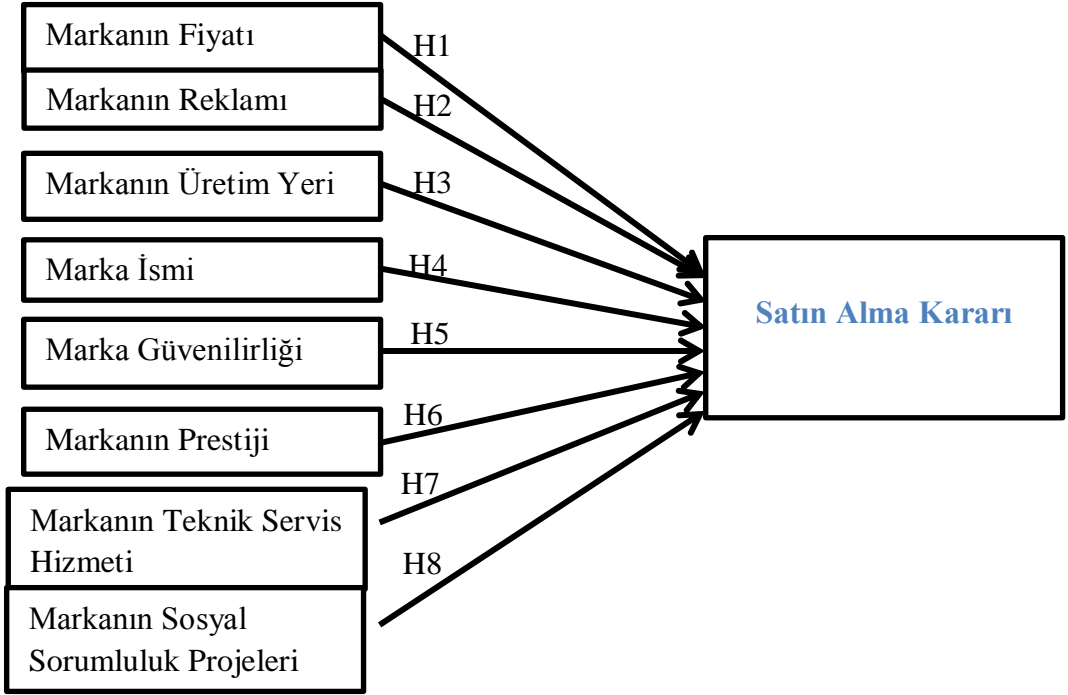

Fiyat, en nihayetinde tüketicinin değerli olduğunu bildiği ürün veya hizmetlerle ilgili değişim yaparken ödeme yapmaya da istekli olduğu bir parasal değerdir. Ödenecek bu paranın tutarı çeşitli şekillerde değişmektedir (Kotler ve Keller, 2012). Tüketiciler ürünü satın almak istediğinde, bir takım makul fiyat aralığı belirlenmektedir. Ürünün mevcut fiyatı, makul fiyat aralığından büyük olduğunda satın alma niyeti zayıflama eğilimindedir (İbrahim vd., 2013). Özellikle cep telefon sahipleri satın almadan önce fiyatlarını karşılaştırmakta ve kararlarını ona göre vermektedir (Çakır ve Demir, 2014).

**H$_1$.** Markanın fiyatı Y kuşağı tüketicilerin cep telefonu satın alma kararını pozitif yönde ve anlamlı olarak etkilemektedir.

Günümüz zorlu medya çevresinde bile ürün reklamları satışların gelişmesine yol açabilmektedir (Kotler ve Armstrong, 2007). Sürekli bir değişimin olduğu cep telefonu pazarı için reklamlar, farklı ve yenilikçi bir durumda tüketicilere mesaj olarak verilmektedir. Cep telefonları reklamlarındaki bu farklılık, tüketicileri daha çok etkilemektedir. Bunun sonucunda cep telefon satın alma kararını da etkilemesi muhtemeldir (Çakır ve Demir, 2014).

**H$_2$.** Markanın reklamı Y kuşağı tüketicilerin cep telefonu satın alma kararını pozitif yönde ve anlamlı olarak etkilemektedir.

Satın alma kararı vermeden önce tüketicilerin üzerinde durduğu bir faktör olan üretim yeri üründen ürüne farklılık göstermektedir. Bu durum özellikle gıda maddeleri için daha belirgindir (Sütütemiz vd., 2009). Tüketici birçok faktörün yanında üretim yeri unsurunu da çoğu zaman göz önüne almaktadır (Tayfun ve Burak, 2016)

**H$_3$.** Markanın üretim yeri Y kuşağı tüketicilerin cep telefonu satın alma kararını pozitif yönde ve anlamlı olarak etkilemektedir.

Bir cep telefonu satın alınmasına yönelik tüm gayretler bu ürünün markasını ve bu markanın ismini ön plana çıkarmaktadır (Chow vd., 2012). Cep telefonları yüksek bir teknoloji aracı olduğundan içindeki donanım ve yazılımı ile marka bütünleşmektedir. Markası bilinen bir cep telefonun donanımı, tüketici zihninde markayı çağrıştırmaktadır. (Lay-Yee vd., 2013).

**H₄.** Markanın ismi Y kuşağı tüketicilerin cep telefonu satın alma kararını pozitif yönde ve anlamlı olarak etkilemektedir.

Cep telefonu markası tercihinde güvenilirlik dikkate alınan bir unsurdur. Bunun nedeni markanın tüketiciye verdiği mesaj ve kalite algısından kaynaklanmaktadır. Güvenilir bir markadan alınacak telefon satın alma kararına da etki edebilir (Marangoz, 2007). Bu doğrultuda marka güvenilirliği ölçüsünde tüketicinin satın alma kararını kolaylaştırır. Tüketicilere yararlı olabilecek yeni ürünlere dikkat çeker (Gavcar ve Didin, 2012).

**H₅.** Markanın güvenilirliği Y kuşağı tüketicilerin cep telefonu satın alma kararını pozitif yönde ve anlamlı olarak etkilemektedir.

Tüketici prestijli bir markanın ürününün kendine yarar sağlayacağını öngörmektedir. Bu durum karşılık bir fayda sağlayacak ardından marka prestij oluşturacak ve markaların kendi içlerinde birbirlerinden farklılaşması sağlanacaktır. Tüketici nitelikli olan markayı kullandıkça bir prestij elde edecektir (Becker ve Murhpy, 1993). Bu yönüyle marka prestiji tüketicinin satın alma davranışlarına önemli ölçüde etki edebilmektedir (Ackerberg, 2001).

**H₆.** Markanın prestiji Y kuşağı tüketicilerin cep telefonu satın alma kararını pozitif yönde ve anlamlı olarak etkilemektedir.

Tüketicilerin marka ile ilgili verdikleri kararlarda teknik servis belirgin bir unsurdur. Bu unsur karar verme ve satış sonrası aşamalarında ön plana çıkmaktadır. Çünkü satın alınan bir ürünün yetkili servisinden alınacak hizmetin kalitesi o markanın gücünü göstermektedir (Cankurt ve Miran, 2010).

**H₇.** Markanın teknik servis hizmetleri Y kuşağı tüketicilerin cep telefonu satın alma kararını pozitif yönde ve anlamlı olarak etkilemektedir.

İşletmenin sosyal olarak sorumlu olduğu yönünde bir imaj yaratmak istemesi, sosyal pazarlamayı ve uygulamalarını gerekli kılmaktadır. Bireyler bilinçlendikçe ve sosyal sorumlulukları artıkça, bu sorumluluklarını davranışlarına yansıtmak için daha fazla fırsat arayışına girmektedirler. Bu noktada en iyi kaynak, sosyal sorumluluklarını tüketiciler ile paylaşan, bu sorumluluğa tüketicileri dâhil eden işletmeler olmaktadır (Demirgüneş, 2015).

**H₈.** Markanın sosyal sorumluluk projeleri satın alma kararını pozitif yönde ve anlamlı olarak etkilemektedir.

## 4. YÖNTEM

Araştırmada cep telefonu satın alma kararında markanın etkisini ölçmek için Sata (2013), Çakır ve Demir (2014) ve Uddin vd. (2014) tarafından yapılan çalışmalardan yararlanılmıştır. Araştırma konusuna yönelik anket formu iki bölümden oluşmaktadır. Anketin birinci bölümünde katılımcılara yönelik demografik özellikler yönelik sorular, ikinci bölümünde ise markanın hangi özelliğinin cep telefonu satın alma kararına etki ettiğini belirlemeye yönelik sorular bulunmaktadır. İfadelerin tümü 5'li likert tipi soru formatında sorulmuştur. Anketler Şanlıurfa Harran Üniversitesi Meslek Yüksek Okullarında görevli genç akademisyenler ve üniversite öğrencileri ile gerçekleştirilmiştir. Araştırmada olasılığa dayalı olmayan örneklem tekniklerden kolayda örnekleme tekniği uygulanmış ve 231 kişiye yüz yüze anket yapılmıştır. Toplanan veriler yapısal eşitlik modellemesi kullanılarak, AMOS ve SPSS istatistiksel paket programları ile analiz edilmiştir.

## 5. BULGULAR

Çalışmada örneklemden toplanan verilerin normal dağılım sergileyip sergilemediği incelenmiş, bu doğrultuda verilerin çarpıklık ve basıklık değerleri analiz edilmiştir. Yapılan analizler neticesinde veriler normal dağılım göstermiştir. Bunun yanında ölçeği oluşturan ifadelerin kendi aralarında tutarlılık gösterip göstermediğini belirlemek amacıyla ölçeğe Cronbach's alpha güvenilirlik analizi uygulanmıştır. Analiz sonucuna göre ölçeğin güvenilirliği 0,81 olarak tespit edilmiştir.

Tablo 1 de katılımcıların demografik özelliklerine göre dağılımları verilmiştir. Buna göre katılımcıların %68'i kadınlardan oluşmakta ve %40,7'si 21-24 yaş aralığındadır. Ayrıca katılımcıların %85,3'ü bekardır ve %38,1'inin aylık geliri 1500 TL'nin üzerindedir.

**Tablo 1:** Demografik veriler

|  | Frekans | Yüzde |
|---|---|---|
| **Cinsiyet** | | |
| Erkek | 74 | 32 |
| Kadın | 157 | 68 |
| **Yaş** | | |
| 18-20 | 76 | 32,9 |
| 21-24 | 94 | 40,7 |
| 25-28 | 21 | 9,1 |
| 29-33 | 10 | 4,3 |
| 34+ | 30 | 13 |
| **Medeni Durum** | | |
| Bekar | 197 | 85,3 |
| Evli | 34 | 14,7 |
| **Aylık Gelir (TL)** | | |
| 0-499 | 16 | 6,9 |
| 500-999 | 73 | 31,6 |
| 1000-1499 | 54 | 23,4 |
| 1500+ | 88 | 38,1 |

Yapılan modifikasyonlar sonucu elde edilen uyum iyiliği değerleri Tablo 2'de gösterilmiştir. Kurulan yapısal model uyum iyiliği değerlerini sağlamaktadır. Ölçeğin uyum indeksleri incelendiğinde: ki-kare istatistiğinin serbestlik derecesine oranı ($\chi^2$/sd) 3,52; iyilik uyum indeksi (GFI) 0,93; karşılaştırmalı uyum indeks (CFI) değeri 0,94 ve yaklaşık hataların ortalama karekökü (RMSEA) 0,06 olarak bulunmuştur.

**Tablo 2:** Yapısal Eşitlik Modeli Uyum İyiliği Değerleri

| | chi2/df | GFI | CFI | NFI | RMSEA |
|---|---|---|---|---|---|
| Önerilen Değerler | <5.0 | ≥0.9 | ≥0.9 | ≥0.9 | <0.08 |
| Mevcut Değer | 3.52 | 0.93 | 0.94 | 0.91 | 0.06 |

**Tablo 3:** Yapısal Eşitlik Modeli Regresyon Ağırlıkları

| **Test Edilen Yol** | **Hipotez** | **Tahmin** | **Anlamlılık** |
|---|---|---|---|
| Markanın Fiyatı → Satın Alma Kararı | H1 | 0,29 | **0,02** |
| Markanın Reklamı → Satın Alma Kararı | H2 | 0,18 | **0,03** |
| Markanın Üretim Yeri → Satın Alma Kararı | H3 | | 0,22 |
| Marka İsmi → Satın Alma Kararı | H4 | 0,45 | **0,01** |
| Marka Güvenilirliği → Satın Alma Kararı | H5 | 0,38 | **0,02** |
| Markanın Prestiji → Satın Alma Kararı | H6 | 0,41 | **0,01** |
| Markanın Teknik Servis Hizmeti → Satın Alma Kararı | H7 | 0,15 | **0,04** |
| Markanın Sosyal Sorumluluk Projeleri → Satın Alma Kararı | H8 | | 0,18 |

Tablo 3'de gösterilen yapısal eşitlik modeli analiz sonuçlarına göre; markanın fiyatı (p=0,02), markanın reklamı (p=0,03), markanın ismi (p=0,01), markanın güvenilirliği (p=0,02), markanın prestiji (p=0,01) ve markanın teknik servis hizmetleri (p=0,04) Y kuşağı tüketicilerin cep telefonu satın alma kararını pozitif yönde anlamlı olarak etki etmektedir. Ancak markanın üretim yeri (p=0,22) ve markanın sosyal sorumluluk projelerine verdiği önem (p=0,18) Y kuşağı tüketicilerin cep telefonu satın alma kararını etki etmemektedir. Bu sonuçlara göre H1, H2, H4, H5, H6, H7 hipotezleri kabul edilmiş ancak H3 ve H8 hipotezleri reddedilmiştir.

## 6.SONUÇ

Bilgi ve iletişim teknolojilerindeki gelişmeler, insanların teknolojiyi kullanma biçimini sürekli olarak değiştirmektedir. Günümüzde tüketicilerin hayatında en yaygın olarak kullanılan teknolojik cihaz ise cep telefonlarıdır. Tüketiciler cep telefonu satın alırken çeşitli markalar arasında bir seçim yapmaktadır. Bu seçimi markaların hangi özelliklerine göre yaptıkları bu araştırmanın çıkış noktasıdır. Markaların çeşitli özellikleri gittikçe birbirine daha benzer hale gelmiş olsa da, tüketiciler markalar arasında kolayca ayrım yapmaktadır. Markalar ise küreselleşen pazar ve rekabetçi piyasa koşullarında tüketici tercihlerini kazanmak ve onlarla uzun süreli bir ilişki kurmak için ürünlerini ön plana çıkaracak hamleler yapmaktadır. Markalar cep telefonları ile ilgili tüketici tercihlerini anlamak amacıyla, uygun bir marka stratejisi oluşturmak zorundadır. Markalar bu stratejiyi prestij, güvenilirlik ve teknik servis hizmetleri gibi birçok özellik ile belirleyebilmektedir.

Analizler neticesinde cep telefonu markasının fiyatı Y kuşağı tüketicilerin cep telefonu satın alma kararını pozitif yönde ve anlamlı olarak etkilemiştir. Bu durum fiyat değişkenin marka seçiminde belirleyici olduğunu göstermektedir. Cep telefonu seçiminde fiyat, geliri kısıtlı olan genç tüketiciler için daha da önemlidir. Bunun yanında fiyat, farklı pazarlama faaliyetleri ile beraber kullanıldığı takdirde başarılı olma şansı daha yüksektir. Ayrıca fiyatlarda yapılan indirimler pazarlamada oldukça sık kullanılan bir yöntemdir. Diğer taraftan fiyat işletmeler tarafından yeni tüketicilerin markayı kullanmasını teşvik eder yâda mevcut tüketicilerin ise daha sıkı tutunmasını sağlar. Markalar cep telefonu pazarında rakiplerinden bir adım daha önde rekabete başlayabilmek için fiyat gibi önemli bir pazarlama faaliyetine gerekenden daha fazla önem vermelidir. Bu doğrultuda markalar cep telefonu fiyatlarını genç kuşağa göre de planlamalıdır. Bunun yanında markanın reklamı Y kuşağı tüketicilerin cep telefonu satın alma kararını pozitif yönde ve anlamlı olarak etkilemiştir. Cep telefonu markalarının reklâm stratejilerine daha fazla önem vermeleri, reklamlarda genç tüketicileri de odak noktaya almaları, rakiplerinden daha fazla reklam harcaması yapmaları ve bu reklamları farklı reklam ortamlarında kullanmaları faydalı olacaktır.

Araştırmada markanın ismi, prestiji ve güvenilirliği Y kuşağı tüketicilerin cep telefonu satın alma kararını pozitif yönde ve anlamlı olarak etkilemiştir. Bunun nedeni ismi ön plana çıkmış prestijli markaların genç tüketicilerin zihninde diğer markalara göre daha güvenilir olmasından kaynaklanabilir. Çünkü prestiji yüksek olan markalar güvenilir olarak kabul edildiği için tüketiciler tarafından daha hızlı kabul edilir ve satın alma aşamasında çok fazla araştırılma gereği duymazlar. Bu yüzden cep telefonu tercihinde genç tüketicileri etkilemek isteyen markaların piyasada öncelikle prestij elde etmeye çalışması onların satın alma kararını hızlandıracak ve markalarına etkisi olumlu olacaktır.

Son olarak araştırmada cep telefonu markalarının teknik servis hizmeti Y kuşağı tüketicilerin cep telefonu satın alma kararını pozitif yönde ve anlamlı olarak etkilemiştir. Bu durumda markaların teknik servis hizmetine erişimi kolaylaştırmaları ve yaygınlaştırmaları doğru bir karar olacaktır. Teknik servis hizmetini ön plana alan cep telefonu markaları Y kuşağı genç tüketicilere daha kolay erişecektir.